\documentclass[a4paper, 12pt]{article}
\begin{document}
\begin{center}
{\Large\textbf{Equivalence of the Self--Dual Model and \\
Maxwell--Chern--Simons Theory on Arbitrary \\
\vspace {2.1mm} 
Manifolds}} \\ 
\vspace {8mm}
\renewcommand{\thefootnote}{$\ast$}
{\large Emil M. Prodanov\footnote {Supported by FORBAIRT scientific research
program and Trinity College, Dublin.} and Siddhartha Sen} \\
\vspace {2mm}
{\it School of Mathematics, Trinity College, Dublin 2, Ireland, \\
e-mail: \hskip 2pt prodanov@maths.tcd.ie, \hskip 2pt sen@maths.tcd.ie}
\end{center}
\vspace{4mm}
\begin{abstract}
Using a group-invariant version of the Faddeev--Popov method we explicitly 
obtain the partition functions of the Self--Dual Model and 
Maxwell--Chern--Simons theory. We show that their ratio  
coincides with the partition function of abelian Chern--Simons theory
to within a phase factor depending on the geometrical properties of the 
manifold. 
\end{abstract}
\scriptsize
%\vskip8.5cm
\normalsize 
\newpage
\section{Introduction}
In three dimensions it is possible to add a gauge-invariant Chern--Simons term
to the Maxwell gauge field action~\cite{jate},~\cite{dejate},~\cite{schon}. 
The resulting Maxwell--Chern--Simons theory has been analyzed completely and
in~\cite{dejate} the entire subject of topologically massive three--dimensional
gauge theories has been set up. Further Maxwell--Chern--Simons theory has been 
used as an effective theory for different models, such as fractional Hall 
effect and high-temperature superconductivity~\cite{wil},~\cite{bala}.

The Self--Dual Model was first studied in detail by
Deser et al.~\cite{dejate} and it was shown in~\cite{deja} that the Self--Dual
Model is equivalent, modulo global differences, to the Maxwell--Chern--Simons 
theory. 

Subsequently, this equivalence has been studied by many authors using variety
of techniques: in the context of bosonisation and at the quantum level (using
Legendre transformation) in the abelian and non-abelian case 
in~\cite{frascha},~\cite{braframascha}; by constraint analysis in~\cite{arias} 
and~\cite{banerr}; by means of Batalin--Fradkin--Tyutin formalism~\cite{baner}; 
in the context of duality~\cite{kog},~\cite{legu},~\cite{steph} and many others.

We address the equivalence with differential geometric tools. It allows us to
reveal global features of these models which, so far, have been overlooked.
We pay particular attention to the zero modes present in the problem. These 
zero modes contain topological information regarding the manifold. By neglecting 
them, i.e. absorbing the divergence due to the zero modes in the normalization
constant, this information is lost. A method, due to Schwarz~\cite{s1}, 
involving invariant integration, allows us to formally consider a key part of 
the zero mode sector from the divergent term. This is enough, as we show, to get
topological information regarding the manifold.

We show that, subject to choice of appropriate renormalizations, the ratio of
the partition functions of the two theories in the presence of currents
is given, modulo phase factor, by the partition function of abelian
Chern--Simons theory with currents. This phase factor captures the geometrical
properties of the manifold. The partition function of Chern--Simons theory
contains a phase factor which captures the topological properties
of the currents (their linking number) and modulo this phase factor it is a
topological invariant (the Ray--Singer torsion of the manifold). Therefore
the Self--Dual Model and Maxwell--Chern--Simons theory are equivalent to
within a phase factor which contains geometrical information about the
manifold and another phase factor which contains information about
the topological properties of the currents. In our considerations we have used 
zeta--function regularised determinants which lead to phase ambiguities.
%\newpage
\section{Review of Schwarz's group-invariant \\
description of the Faddeev--Popov method}
\setcounter{equation}{0}
In this section for the sake of completeness we will review,  
following~\cite{s1}, the method of reducing an integral of a function 
with some symmetries over some space to an integral over a lower-dimensional 
space. \\
Take $M$ to be a Riemannian manifold and $G$ --- a compact group.
Let $W=\raise 2pt \hbox{$M$}  /  \lower 2pt \hbox{$G$}$ denote the space
of orbits.\\
Using the Riemannian metric, there are no problems in defining volume elements 
on $W$ and $M$. \\
Let $\lambda(x)$ be the volume of the orbit $Gx$ with respect to the Riemannian
metric on $M$.          
$\lambda(x)$ is $G$-invariant (since  $\lambda(gx)= \lambda(x)$ for $g\in G$).
Therefore $\lambda(x)$ is a function on $W=\raise 2pt \hbox{$M$}  /  \lower 2pt
\hbox{$G$}$.\\          
Let $f(x)$ be $G$-invariant function on $W=\raise 2pt \hbox{$M$}  /  \lower 2pt
\hbox{$G$}$.\\  
Hence:
\begin{equation}
\int\limits_M \ f(x) d\mu = \int\limits_{\raise 1pt \hbox{$\scriptstyle M$}
/  \lower 1pt\hbox{$\scriptstyle G$}} \ f(x)\lambda(x) dv 
\end{equation}
Define the linear operator $T_x: Lie(G) \longrightarrow T_x(M)$, where $T_x(M)$
is the tangent space to $M$ at $x$ and $Lie(G)$ is the Lie algebra of the 
group $G$.\\
Let $H_x$ be the stabilizer of the group $G$ at $x$, i.e.:
\begin{equation}
H_x x=x
\end{equation}
Therefore:
\begin{equation}
ker(T_x) = Lie(H_x) = \mathcal{H}_x 
\end{equation}
Consider the linear operator $\widetilde {T_x}: \raise 4pt\hbox{$\ Lie(G)$}
\!\! \Bigm/ \!\!\lower 4pt \hbox{$\ Lie(H_x)$} \quad \longrightarrow
\quad T_x(M)$.\\
The operator $T^\dagger_x T_x^{\phantom{\dagger}}$ is non-degenerate if, and
only if, $G$ acts with discrete stabilizers.
The operator $\widetilde {T^\dagger_x} \widetilde{T_x^{\phantom{\dagger}}}$ is
always non-degenerate.
The quotient  $\raise 2pt \hbox{$G$}  /  \lower 2pt \hbox{$H_x$}$ is 
homeomorphic to the orbit $Gx$ under the map $g \mapsto gx$ for $g \in G$. 
The differential of this map at the identity coincides with the operator  
$\widetilde {T_x}$.\\
Therefore:
\begin{equation}
vol(Gx)=vol\bigl(\!\!\raise 4pt \hbox{$\ G$} \! / \!\! \!\lower 4pt\hbox
{$\ H_x$}\bigr) \vert det \widetilde {T_x} \vert =
vol\bigl(\!\!\raise 4pt \hbox{$\ G$} \! / \!\!\!\lower 4pt\hbox
{$\ H_x$}\bigr) (det \widetilde {T^\dagger_x}
\widetilde{T_x^{\phantom{\dagger}}}) ^  {\scriptscriptstyle {1/2}}  
\end{equation}
But
\begin{equation}
vol(G) = \int\limits_G \ Dg = \int\limits_{\raise 1pt \hbox{$
\scriptstyle G$}  / \lower 1pt\hbox{$\scriptstyle H_x$}} D[g] \hskip 4pt
vol(H_x) = vol(H_x) vol\bigl(\!\!\raise 4pt \hbox{$\ G$} \! / \!\! \!\lower 4pt
\hbox{$\ H_x$}\bigr)
\end{equation}
Take $vol(G)$ to be normalized to 1.\\
Then the volume of the orbit of the group is:
\begin{equation}
\lambda(x)=\frac {1}{vol(H_x)} (det \widetilde {T^\dagger_x}
\widetilde{T_x^{\phantom{\dagger}}}) ^  {\scriptscriptstyle {1/2}}
\end{equation}
We now assume that all stabilizers are conjugate 
and have the same volume $vol(H)$. \\
Then:
\begin{equation}
\int\limits_M \ f(x) d\mu = \frac {1}{vol(H)} \int\limits_{\raise 1pt
\hbox{$\scriptstyle M$}  / \lower 1pt\hbox{$\scriptstyle G$}} f(x)
(det \widetilde {T^\dagger_x}\widetilde{T_x^{\phantom{\dagger}}})
^ {\scriptscriptstyle {1/2}} dv
\end{equation}
With this formula we have restricted the gauge freedom by picking up only one 
representative from each orbit. Alternatively, we could have imposed a
gauge-fixing condition and inserted it in the action together with 
the Faddeev--Popov determinant. This would bring a delta-function of the 
gauge-fixing condition into the integrand and therefore would define a 
subspace in M. If this gauge-fixing condition is appropriate, this subspace 
would intersect each orbit exactly once and therefore the integration would 
pick up one representative of each orbit. For our further considerations 
(quadratic action functionals) we can also choose the resolvent 
method~\cite{s2} --- an invariant and beautiful form of the 
Faddeev--Popov trick.

\section{Partition Function of the Self--Dual Model}
The Self-Dual Model is given by the action:
\begin{equation}
S_{\scriptscriptstyle SD}=\int\limits_{M}(f_\mu f^{\mu} +
\epsilon_{\mu\nu\lambda}f^{\mu}\partial^{\nu}f^{\lambda}) 
d^{\scriptscriptstyle 3}x 
\quad = \quad
\Bigl\langle f, (\mathrm{1\mkern-5muI\!}\hskip2pt + \ast \hskip 2pt
d_{\scriptscriptstyle 1})f \Bigr\rangle
\end{equation}
where $d_{\scriptscriptstyle p}$ is the map from the space of all p-forms
to the space of (p+1)-forms, i.e. 
$d_{\scriptscriptstyle p}: \hskip4pt 
\Omega^{\scriptstyle p} (M) \longrightarrow
\Omega^{\scriptstyle p+1}(M)$, and $\ast$ is the Hodge star operator:
$\ast: \hskip4pt \Omega^{\scriptstyle p} (M) \longrightarrow 
\Omega^{\scriptstyle m-p} \hskip 8pt (m = dim M = 3)$. The Hodge star operator
explicitly depends on the metric of the manifold $M$. \\
Here we have introduced the notation of differential forms. Namely, we can 
write the scalar product of two p-forms $a_{\scriptscriptstyle p}$ and 
$b_{\scriptscriptstyle p}$ as: \hskip 4pt 
$\langle a_{\scriptscriptstyle p}, b_{\scriptscriptstyle p} \rangle = 
\int\limits_{M} a_{\scriptscriptstyle p} \wedge \ast 
b_{\scriptscriptstyle p}$. \\  
In this notation $\int\limits_{M}f_\mu f^{\mu}d^{\scriptscriptstyle 3}x$ 
may be written as $\int\limits_{M} f \wedge \ast f$ and thus 
explicitly involves the metric and violates 
topological invariance. By ``topological invariance'' we will always mean 
metric-independence. The Chern--Simons term, in spite of the presence of 
$\ast$, is actually a topological invariant as it can be written simply as
$\int\limits_{\scriptscriptstyle M} A \wedge dA$ involving only  
differential forms which do not depend on the metric.
The partition function of the model is:
\begin{equation}
Z_{\scriptscriptstyle SD} = \frac{1}{N_{\scriptscriptstyle SD}}
\int\limits_{\Omega^{\scriptscriptstyle 1}(M)} \mathcal {D}
f \hskip 4pt e^{-i \hskip3pt \bigl\langle f, \hskip4pt 
(\mathrm{1\mkern-5muI\!}\hskip2pt + 
\ast d_{\scriptscriptstyle 1})f \bigr\rangle}
\end{equation} 
where $\Omega^{\scriptscriptstyle 1}(M)$ is the space of all 1-forms on $M$ 
and $N_{\scriptscriptstyle SD}$ is a normalization factor.\\
The operator $\mathrm{1\mkern-5muI\!}\hskip2pt + \ast \hskip 2pt
d_{\scriptscriptstyle 1}$ is self-adjoint. \\
Now we will extract the zero-mode dependence from the action functional. To do
so, decompose $\Omega^{\scriptscriptstyle 1}(M)$:
\begin{equation}
\Omega^{\scriptscriptstyle 1}(M)=ker(\mathrm{1\mkern-5muI\!}\hskip2pt + \ast
\hskip 2pt d_{\scriptscriptstyle 1}) \oplus 
ker(\mathrm{1\mkern-5muI\!}\hskip2pt + \ast\hskip 2pt d_{\scriptscriptstyle 1})
^{\perp}  
\end{equation} 
Therefore:
\begin{equation} 
Z_{\scriptscriptstyle SD} = \frac{1}{N_{\scriptscriptstyle SD}}
vol\Bigl(ker(\mathrm{1\mkern-5muI\!}\hskip2pt + 
\ast\hskip 2pt d_{\scriptscriptstyle 1})\Bigr) 
det'\Bigl(i (\mathrm{1\mkern-5muI\!}\hskip2pt + 
\ast\hskip 2pt d_{\scriptscriptstyle 1})\Bigr)^{\scriptscriptstyle -1/2} 
\end{equation} 
Here we have retained, formally, the infinite volume factor. This can be
removed by choosing the normalization constant $N_{\scriptscriptstyle SD}$
appropriately. \\
Witten has shown~\cite{w} how to deal with $i$ in $det'(iT)$ for 
some operator $T$ using $\zeta$--regularization technique. He found that 
$i$ leads to a phase factor, depending on the $\eta$--function of 
the operator $T$ and explicitly involving the metric of the manifold. 
For our case we have:
\begin{equation}
det'\Bigl(i (\mathrm{1\mkern-5muI\!}\hskip2pt + 
\ast\hskip 2pt d_{\scriptscriptstyle 1})\Bigr)^{\scriptscriptstyle -1/2} = 
e^{- \frac{i \pi}{4} \eta\bigl(0, \hskip4pt (\mathrm{1\mkern-5muI\!}\hskip2pt +
\ast\hskip 2pt d_{\scriptscriptstyle 1})\bigr)}
det'(\mathrm{1\mkern-5muI\!}\hskip2pt +
\ast\hskip 2pt d_{\scriptscriptstyle 1})^{\scriptscriptstyle -1/2}
\end{equation}
Finally, the partition function of the Self--Dual Model is:
\begin{equation}
Z_{\scriptscriptstyle SD} = \frac{1}{N_{\scriptscriptstyle SD}}
e^{- \frac{i \pi}{4} \eta\bigl(0, \hskip4pt (\mathrm{1\mkern-5muI\!}\hskip2pt +
\ast\hskip 2pt d_{\scriptscriptstyle 1})\bigr)}
vol\Bigl(ker(\mathrm{1\mkern-5muI\!}\hskip2pt + 
\ast\hskip 2pt d_{\scriptscriptstyle 1})\Bigr) 
det'(\mathrm{1\mkern-5muI\!}\hskip2pt +
\ast\hskip 2pt d_{\scriptscriptstyle 1})^{\scriptscriptstyle -1/2}
\end{equation}
\section{Partition Function of \\ Maxwell--Chern--Simons Theory}
The action of Maxwell--Chern--Simons Theory is:
\begin {eqnarray}
S_{\scriptscriptstyle MCS} &=& \int\limits_{M} (F_{\mu\nu} F^{\mu\nu} + 
\epsilon_{\mu\nu\lambda}
A^{\mu} \partial^{\nu} A^{\lambda})d^{\scriptscriptstyle 3}x  \nonumber \\
&& \mkern-29mu 
= \quad
\langle A, \hskip2pt d_{\scriptscriptstyle 1}^{\dagger}d_{\scriptscriptstyle 1}^
{\phantom{\dagger}}A \rangle + \langle A, \hskip2pt \ast d_{\scriptscriptstyle 1}A \rangle
\end{eqnarray}
where $d_{\scriptscriptstyle k}^{\dagger}: \Omega^{\scriptscriptstyle k+1}(M) 
\longrightarrow  \Omega^{\scriptscriptstyle k}(M).$\\
In this case the topological invariance is explicitly violated by the Maxwell 
term. \\
On three-dimensional manifolds we have: $d_{\scriptscriptstyle 1}^{\dagger}=
\ast \hskip 4pt  d_{\scriptscriptstyle 1}\ast.$
Therefore we may write the partition function as:
\begin{equation}
Z_{\scriptscriptstyle MCS} = \frac{1}{N_{\scriptscriptstyle MCS}}
\int\limits_{\Omega^{\scriptscriptstyle 1}(M)}
\mathcal{D}A \hskip 4pt e^{-i \bigl\langle A, \hskip 4pt 
\left(\ast d_{\scriptscriptstyle 1} +
(\ast d_{\scriptscriptstyle 1})^{\scriptscriptstyle 2} \right)A \bigr\rangle}
\end{equation}
The operator $\ast d_{\scriptscriptstyle 1} +
(\ast d_{\scriptscriptstyle 1})^{\scriptscriptstyle 2}$ is self-adjoint.\\
We proceed to explicitly calculate the partition function.
The theory has a gauge invariance under
gauge transformations \hskip 4pt $A_{\mu} \longrightarrow A_{\mu} -
\partial_{\mu}\lambda$ , \hskip 4pt i.e.:
\begin{equation} 
A \longrightarrow A \hskip 6pt + \hskip 6pt d_{0} \hskip2pt 
\Omega^{\scriptscriptstyle 0}(M)
\end{equation}
We thus have an infinite-dimensional analogue of the situation reviewed in
Section 2.
To proceed we pick up one representative of each equivalence class $[A]$, 
where $[A]= {A + d_{\scriptscriptstyle 0} \hskip2pt \Omega^
{\scriptscriptstyle 0}(M)}.$ \hskip4pt
To do this we impose the gauge condition $\partial_{\mu} A^{\mu} = 0$, 
that is $d_{\scriptscriptstyle 0}^{\dagger}\hskip 2pt A = 0$.\\
This ensures that the space of orbits of the gauge group in the 
space of all 1-forms is orthogonal to the space of those $A$'s, for which 
$d_{\scriptscriptstyle 0}^{\dagger}A=0$ and so we will pick up only one 
representative of each orbit.\\
Then the operator $d_{\scriptscriptstyle 0}$ plays the role of $\widetilde 
{T^\dagger_x}$ of Section 2, i.e. the 
stabilizer consists of those elements of  
$\Omega^{\scriptscriptstyle 0}(M)$, for which
$d_{\scriptscriptstyle 0} \hskip2pt 
\Omega^{\scriptscriptstyle 0}(M)=0$ (the constant functions). Hence: 
$H=\mathrm{I\!R\!}$.\\
Therefore:
\begin{equation}
Z_{\scriptscriptstyle MCS} = \frac{1}{N_{\scriptscriptstyle MCS}}
\hskip 3pt \frac{1}{vol(H)} \int\limits_{\raise 2pt\hbox{$\scriptstyle  
\Omega^{\scriptscriptstyle 1}(M)$} / \lower 2pt\hbox{$\scriptstyle G$}} 
\mathcal{D}
A \hskip 4pt e^{-i \hskip4pt
\bigl\langle A,\hskip4pt \left(\ast d_{\scriptscriptstyle 1} + 
(\ast d_{\scriptscriptstyle 1})^2 \right)A \bigr\rangle}
(det'{d^\dagger_{\scriptscriptstyle 0}} 
{d_{\scriptscriptstyle 0}^ {\phantom{\dagger}}}) 
^ {\scriptscriptstyle {1/2}}
\end{equation}
The operator in the exponent has zero-modes.\\
Let $A \in ker \Bigl(\ast d_{\scriptscriptstyle 1} +
(\ast d_{\scriptscriptstyle 1})^{\scriptscriptstyle 2} \Bigr),\hskip 4pt 
$i.e.$ \hskip 4pt
\ast \hskip 2pt d_{\scriptscriptstyle 1}A +
\ast \hskip 2pt d_{\scriptscriptstyle 1}\!
\ast d_{\scriptscriptstyle 1}A = 0.$ \hskip5pt There are two situations to
consider. We can take \hskip 3pt 
$\ast \hskip 2pt d_{\scriptscriptstyle 1}A = 0,$ that is:
$A \in ker(\ast d_{\scriptscriptstyle 1})$, \hskip5pt or \hskip5pt
$A \notin ker(\ast d_{\scriptscriptstyle 1})$, \hskip 6pt i.e.
$\ast \hskip 2pt d_{\scriptscriptstyle 1}A \ne 0$, but  
$\ast \hskip 2pt d_{\scriptscriptstyle 1}A = 
-(\ast \hskip 2pt d_{\scriptscriptstyle 1})^{\scriptscriptstyle 2}A.$
In the second case $\ast \hskip 2pt d_{\scriptscriptstyle 1}$ has inverse 
$(\ast \hskip 2pt d_{\scriptscriptstyle 1})^{\scriptscriptstyle -1}$.
Therefore: \hskip 4pt $A  = - \ast \hskip 2pt d_{\scriptscriptstyle 1}A$,
\hskip 4pt which means that \hskip 4pt $(\mathrm{1\mkern-5muI\!} + 
\ast \hskip 2pt d_{\scriptscriptstyle 1})A = 0,$ \hskip 4pt
i.e. \hskip 4pt $A \in ker(\mathrm{1\mkern-5muI\!} +
\ast \hskip 2pt d_{\scriptscriptstyle 1}).$\\
By definition $ker(\ast \hskip 2pt d_{\scriptscriptstyle 1})
\bigcap ker(\mathrm{1\mkern-5muI\!} + \ast \hskip 2pt d_{\scriptscriptstyle 1}) = \emptyset.$\\
It is easy to see that  $ker(\ast \hskip 2pt d_{\scriptscriptstyle 1})$ and
$ ker(\mathrm{1\mkern-5muI\!} + \ast \hskip 2pt d_{\scriptscriptstyle 1})$  
are orthogonal:\\
Let $f \in  ker(\mathrm{1\mkern-5muI\!} + \ast \hskip 2pt 
d_{\scriptscriptstyle 1})$  and $g \in ker(\ast \hskip 2pt  
d_{\scriptscriptstyle 1}).$\\
$\langle f, \hskip 4pt g \rangle \hskip 4pt = \hskip 4pt \langle f, \hskip 4ptg \rangle + 
\langle f, \hskip 4pt \ast \hskip 2pt 
d_{\scriptscriptstyle 1}g \rangle  \hskip 4pt = \hskip 4pt 
\Bigl\langle f, \hskip 4pt (\mathrm{1\mkern-5muI\!} + 
\ast \hskip 2pt d_{\scriptscriptstyle 1})g\Bigr
\rangle \hskip 4pt = \hskip 4pt $ \\
$= \hskip 4pt \Bigl\langle (\mathrm{1\mkern-5muI\!} +  
\ast \hskip 2pt d_{\scriptscriptstyle 1})f, \hskip 4pt g \Bigr\rangle 
\hskip 4pt = \hskip 4pt 0$, \hskip 4pt since 
$(\mathrm{1\mkern-5muI\!} + \ast \hskip 2ptd_{\scriptscriptstyle 1})$ is
self-adjoint and 
$f \in  ker(\mathrm{1\mkern-5muI\!} + \ast \hskip 2ptd_{\scriptscriptstyle 1}).
\\
$So, $ker(\ast \hskip 2pt d_{\scriptscriptstyle 1})$ is the orthogonal 
complement of
$ ker(\mathrm{1\mkern-5muI\!} + \ast \hskip 2pt d_{\scriptscriptstyle 1}).$\\
Therefore we can write:
\begin{equation} 
ker\Bigl(\ast d_{\scriptscriptstyle 1} + 
(\ast d_{\scriptscriptstyle 1})^{\scriptscriptstyle 2}\Bigr)  = 
ker(\ast d_{\scriptscriptstyle 1}) \oplus ker(\mathrm{1\mkern-5muI\!} + 
\ast \hskip 2ptd_{\scriptscriptstyle 1})
\end{equation}
The partition function is then given by:
\begin{eqnarray}
Z_{\scriptscriptstyle MCS} = \frac{1}{N_{\scriptscriptstyle MCS}}
\frac{1}{vol(H)}  
e^{- \frac{i \pi}{4} \eta\bigl(0, \hskip4pt 
\ast d_{\scriptscriptstyle 1} +  
		(\ast d_{\scriptscriptstyle 1})^{\scriptscriptstyle 2}\bigr)}
vol\biggl(ker\Bigl(\ast d_{\scriptscriptstyle 1} +  
		(\ast d_{\scriptscriptstyle 1})^{\scriptscriptstyle 2}\Bigl)
				\biggr) \nonumber \\
\mkern-50mu
det'\Bigl(\ast d_{\scriptscriptstyle 1} +
 		(\ast d_{\scriptscriptstyle 1})^{\scriptscriptstyle 2}\Bigr)
			^{\scriptscriptstyle {-1/2}}
det'(d_{\scriptscriptstyle 0}^\dagger d_{\scriptscriptstyle 0}^
			{\phantom{\dagger}})^{\scriptscriptstyle {1/2}}
\end{eqnarray} 
Because of (18) we can write this as:
\begin{eqnarray} 
Z_{\scriptscriptstyle MCS} = \frac{1}{N_{\scriptscriptstyle MCS}} \hskip3pt
\frac{1}{vol(H)}
e^{- \frac{i \pi}{4} \eta\bigl(0, \hskip4pt 
\ast d_{\scriptscriptstyle 1} +  
		(\ast d_{\scriptscriptstyle 1})^{\scriptscriptstyle 2}\bigr)} 
vol\Bigl(ker(\ast d_{\scriptscriptstyle 1})\Bigr)
vol\Bigl(ker(\mathrm{1\mkern-5muI\!}\hskip2pt + 
\ast \hskip 2ptd_{\scriptscriptstyle 1})\Bigr)  
\nonumber \\
det'\Bigl(\ast d_{\scriptscriptstyle 1} +
(\ast d_{\scriptscriptstyle 1})^{\scriptscriptstyle 2}\Bigr)
^{\scriptscriptstyle {-1/2}}
det'(d_{\scriptscriptstyle 0}^\dagger d_{\scriptscriptstyle 0}^{\phantom
{\dagger}})^{\scriptscriptstyle {1/2}}   
\end{eqnarray}
Let us now consider $det'\Bigl(\ast d_{\scriptscriptstyle 1} +
(\ast d_{\scriptscriptstyle 1})^{\scriptscriptstyle 2}\Bigr)
^{\scriptscriptstyle {-1/2}}$.\\
In the infinite-dimensional case we have to take into account the 
multiplicative anomaly, i.e. the fact that the determinant of a product of
operators is not always the product of the determinants of the operators. \\ 
For our case we will show that:
\begin{equation}
det'\Bigl(\ast d_{\scriptscriptstyle 1} + (\ast d_{\scriptscriptstyle 1})^
{\scriptscriptstyle 2}\Bigr)=(-1)^{\psi}
det'(\ast d_{\scriptscriptstyle 1})
det'(\mathrm{1\mkern-5muI}+\ast d_{\scriptscriptstyle 1})
\end{equation}
\noindent
where: $\psi=\zeta\biggl(0, -\Bigl(\ast d_{\scriptscriptstyle 1}
+ (\ast d_{\scriptscriptstyle 1})^{\scriptscriptstyle 2}\Bigr)_
{\scriptscriptstyle -}\biggr) -
\zeta\Bigl(0, - (\ast d_{\scriptscriptstyle 1})_{\scriptscriptstyle -}\Bigr)
- \zeta\Bigr(0, -(\mathrm{1\mkern-5muI}+\ast d_{\scriptscriptstyle 1})_
{\scriptscriptstyle -}\Bigr).$ \\
The meaning of $\zeta(0, A_{\scriptscriptstyle -})$ will become clear from the 
context of the proof.\\
Take $A$ to be some operator without zero-modes. \\
We can always write $A$ in the form: $A=\left(\matrix{A_{\scriptscriptstyle +} &
\cr & A_{\scriptscriptstyle -}}
\right), \quad 
\vert A \vert = \left(\matrix{A_{\scriptscriptstyle +} & \cr &
- A_{\scriptscriptstyle -}}\right)$
\noindent
where $A_{\scriptscriptstyle \pm}:\quad  \Gamma_{\scriptscriptstyle \pm}
\longrightarrow \Gamma_{\scriptscriptstyle \pm}$ and $\Gamma_{\scriptscriptstyle
\pm}$ is the space spanned by eigenvectors of $A$ corresponding to 
positive (negative) eigenvalues. The operator $\vert A \vert$ has positive 
eigenvalues only. 
If we denote by $\lambda_{\scriptscriptstyle n}$ the eigenvalues of the 
operator $A$, for $det \vert A \vert$ we can use the $\zeta$-regularised 
expresion to define:
\begin{equation}
det \vert A \vert = \prod\limits_{\scriptscriptstyle n=1}^{\infty}
\lambda_{\scriptscriptstyle n} =
e^{\scriptscriptstyle -\zeta'\left(0, \hskip3pt \vert A \vert\right)}
\end{equation}
For any real number $\beta$ we have:
\begin{equation}
det(\beta\vert A \vert)=\beta^{\scriptscriptstyle -\zeta(0, \vert A \vert)}
det\vert A \vert
\end{equation}
The determinant of any operator $A$ can be written as:
\begin{equation}
det(A) = det(A_{\scriptscriptstyle +})det\Bigl(-(-A_{\scriptscriptstyle -})
\Bigr) = (-1)^{\scriptscriptstyle -\zeta(0, -A_{\scriptscriptstyle -})}
det \vert A \vert 
\end{equation}
Let $A=\ast d_{\scriptscriptstyle 1}, \quad
B=\mathrm{1\mkern-5muI\!}\hskip2pt +\ast\hskip2pt d_{\scriptscriptstyle 1}$.\\
For the determinants we write the formal expressions which are always to be
regularised using (22) as:
\begin{equation}
det\vert A \vert = \prod\limits_{\scriptscriptstyle n=1}^{\infty}
\vert \lambda_{\scriptscriptstyle n} \vert \hskip3pt, \hskip2cm
det\vert B \vert = \prod\limits_{\scriptscriptstyle n=1}^{\infty}
\vert 1+\lambda_{\scriptscriptstyle n} \vert
\end{equation}
Denote by $F(U, V) = \frac{det(UV)}{det(U) det(V)}$ the multiplicative 
anomaly. In general $F(U, V) \ne 1$, as well as: 
$ln\hskip 2pt det (U) \ne tr\hskip2pt ln (U)$.\\
We have:
\begin{equation}
ln F( \vert A \vert, \vert B \vert) =
- \frac{d}{ds}\Bigl(\zeta(s, \vert AB \vert) - \zeta(s, \vert A \vert)
- \zeta(s, \vert B \vert)\Bigr)_{\scriptscriptstyle s=0} = 0
\end{equation}
$\zeta(s, \vert U \vert)$ can be expressed as:
\begin{equation}
\zeta (s, \vert U \vert) = \zeta (s, U_{\scriptscriptstyle +}) +
\zeta (s, -U_{\scriptscriptstyle -})
\end{equation}
Therefore:
\begin{eqnarray}
ln F( \vert A \vert, \vert B \vert)  = 
&& \mkern-25mu\!\!\!\!\! 
- \frac{d}{ds} \Bigl( \zeta \bigl(s, (AB)_{\scriptscriptstyle +} \bigr) +
\zeta \bigl( s, (AB)_{\scriptscriptstyle -} \bigr ) -
\nonumber \\
& - & \zeta(s, A_{\scriptscriptstyle +}) - \zeta(s, A_{\scriptscriptstyle -}) -
\zeta(s, B_{\scriptscriptstyle +}) - \zeta(s, B_{\scriptscriptstyle -})\Bigr)
_{\scriptscriptstyle s=0}
\end{eqnarray}
For all operators entering (28) we can apply the analysis of~\cite{e}. 
This analysis holds for the case of a smooth and compact manifold.
The Seeley--De Witt formula:
\begin{equation}
\zeta(s, U)=\frac{1}{\Gamma(s)}\biggl(\sum\limits_{\scriptscriptstyle n=0}^
{\infty}\frac{A_{\scriptscriptstyle n}}{s+n-\frac{D}{2}} + J(s)\biggr)
\end{equation}
where $A_{\scriptscriptstyle n}$ are the heat-kernel coefficients, $D$ is the
dimension of the manifold and $J(s)$ is some analytic function, leads to the
fact that the multiplicative anomaly will vanish when $D=2$ or $D$ is odd.
Therefore:
\begin{equation}
det'\Bigl(\ast d_{\scriptscriptstyle 1} + (\ast d_{\scriptscriptstyle 1})^
{\scriptscriptstyle 2}\Bigr)=e^{i \pi \psi}
det'(\ast d_{\scriptscriptstyle 1})
det'(\mathrm{1\mkern-5muI}+\ast d_{\scriptscriptstyle 1})
\end{equation}
where:\quad $\psi=\zeta\biggl(0, -\Bigl(\ast d_{\scriptscriptstyle 1}
+ (\ast d_{\scriptscriptstyle 1})^{\scriptscriptstyle 2}\Bigr)_
{\scriptscriptstyle -}\biggr) -
\zeta\Bigl(0, - (\ast d_{\scriptscriptstyle 1})_{\scriptscriptstyle -}\Bigr)
- \zeta\Bigr(0, -(\mathrm{1\mkern-5muI}+\ast d_{\scriptscriptstyle 1})_
{\scriptscriptstyle -}\Bigr)$. 
Note that the appearance of the phase factor is not due to the multiplicative
anomaly. We have used the fact that the multiplicative anomaly vanishes for
the moduli of the operators. However, from (24) we see that we are forced to
include some phase ambiguity which is related to the ``negative'' parts of
the operators --- otherwise we would not be able to define a zeta-function 
regularized expressions for operators which have negative 
eigenvalues\footnote[1]{For other examples of phase ambiguities associated 
with $\zeta$-regularised determinants see for instance~\cite{sen} 
and~\cite{des3}.}. \\
It follows that the partition function of Maxwell--Chern--Simons theory can be
written as:
\begin {eqnarray} 
Z_{\scriptscriptstyle MCS} = 
\frac{e^{- \frac{i \pi}{2}\psi - \frac{i \pi}{4} \eta\bigl(0, \hskip4pt 
\ast d_{\scriptscriptstyle 1} +  
(\ast d_{\scriptscriptstyle 1})^{\scriptscriptstyle 2}\bigr)}}
{N_{\scriptscriptstyle MCS}} \hskip 3pt
\frac{1}{vol(H)}
vol\Bigl(ker(\ast d_{\scriptscriptstyle 1})\Bigr)&&
\mkern-25mu 
vol\Bigl(ker(\mathrm{1\mkern-5muI\!} + 
\!\ast \hskip 2ptd_{\scriptscriptstyle 1})\Bigr) \nonumber \\
&& 
\mkern-250mu
det'(\ast d_{\scriptscriptstyle 1})^{\scriptscriptstyle {-1/2}}
det'(\mathrm{1\mkern-5muI\!}\hskip2pt + 
\ast d_{\scriptscriptstyle 1})^{\scriptscriptstyle {-1/2}} 
det'(d_{\scriptscriptstyle 0}^\dagger d_{\scriptscriptstyle 0}^{\phantom
{\dagger}})^{\scriptscriptstyle {1/2}}
\end {eqnarray}
The $\ast$-operator is invertible, hence: \hskip4pt
$ker(\ast d_{\scriptscriptstyle 1}) = ker(d_{\scriptscriptstyle 1}).$\\
We proceed to write (31) in a more convenient form in which topological
features are highlighted. We note that the projection map 
$ker(d{\scriptscriptstyle k}) \longrightarrow H^
{\scriptstyle q}_{\scriptscriptstyle dR}(M)$ induces the isomorphism:
\begin{equation}
\phi_{\scriptstyle q}: \mathcal{H}^{\scriptstyle q}(M)
\longrightarrow H^{\scriptstyle q}_{\scriptscriptstyle dR}(M),
\end{equation}
where $H^{\scriptstyle q}_{\scriptscriptstyle dR}(M)$ is the $q^{th}$ deRham
cohomology group and $\mathcal{H}^{\scriptstyle q}(M)$ is the space of the
harmonic q-forms. Therefore:
\begin{equation}
vol\Bigl(ker(d_{\scriptscriptstyle q})\Bigr) =
\vert det\phi_{\scriptscriptstyle q}
\vert ^{\scriptscriptstyle -1}
vol\Bigl( H^{\scriptscriptstyle q}_{\scriptscriptstyle dR}(M)\Bigr).
\end{equation}
All inner products are in the space of the harmonic forms. In the deRham
cohomology group there is no $\ast$-operator --- it is purely topological. 
The reason for introducing the deRham cohomology groups is that we would like 
to isolate the metric dependence which is present in the volume elements and in
the inner products. \\
The stabilizer $H$ consists of those elements of $\Omega^{\scriptscriptstyle 0}
(M)$ for which \hskip 3pt 
$d_{\scriptscriptstyle 0}\Omega^{\scriptscriptstyle 0}=0$, \hskip 4pt
i.e. $H=ker(d_{\scriptscriptstyle 0})$. So the volume of the stabilizer is:
\begin{equation}
vol(H) = vol\Bigl(H^{\scriptscriptstyle 0}_{\scriptscriptstyle dR}(M)\Bigr) 
\vert det\phi_{\scriptscriptstyle 0}\vert^{\scriptscriptstyle -1}.
\end{equation}
Finally, the partition function of Maxwell-Chern-Simons theory is:
\begin {eqnarray}
Z_{\scriptscriptstyle MCS}  = 
\frac{e^{- \frac{i \pi}{2} \psi - \frac{i \pi}{4} \eta\bigl(0, \hskip4pt 
\ast d_{\scriptscriptstyle 1} +  
(\ast d_{\scriptscriptstyle 1})^{\scriptscriptstyle 2}\bigr)}}
{N_{\scriptscriptstyle MCS}} \hskip 3pt
\frac {vol\Bigl( H^{\scriptscriptstyle 1}_{\scriptscriptstyle dR}(M)\Bigr)}
{vol\Bigl( H^{\scriptscriptstyle 0}_{\scriptscriptstyle dR}(M)\Bigr)}
&&\mkern-30mu
\frac {det(\phi^{\dagger}_{\scriptscriptstyle 0}
\phi^{\phantom{\dagger}}_{\scriptscriptstyle 0})^{\scriptscriptstyle 1/2}}
{det(\phi^{\dagger}_{\scriptscriptstyle 1}
\phi^{\phantom{\dagger}}_{\scriptscriptstyle 1})^{\scriptscriptstyle 1/2}}
det'(d^{\dagger}_{\scriptscriptstyle 0}
d^{\phantom{\dagger}}_{\scriptscriptstyle 0})^{\scriptscriptstyle 1/2}
\nonumber \\
&& \mkern-250mu
det'(\ast d_{\scriptscriptstyle 1})^{\scriptscriptstyle {-1/2}}  
det'(\mathrm{1\mkern-5muI\!} + \!\ast \hskip 2pt d_{\scriptscriptstyle 1})
^{\scriptscriptstyle {-1/2}}
vol\Bigl(ker(\mathrm{1\mkern-5muI\!} + 
\!\ast \hskip 2ptd_{\scriptscriptstyle 1})\Bigr)
\end{eqnarray}
The volume elements in (35) are not finite. We can define the normalization
constant to be the inverse of the ratio of the volumes of the deRham cohomology
groups. Using this fact and (13), (20), (30), and (33) we get: 
\begin{eqnarray}
Z_{\scriptscriptstyle MCS}  =  
e^{- \frac{i \pi}{2} \psi  - \frac{i \pi}{4} \eta\bigl(0, \hskip4pt 
\ast d_{\scriptscriptstyle 1} +  
(\ast d_{\scriptscriptstyle 1})^{\scriptscriptstyle 2}\bigr) +
\frac{i \pi}{4} \eta(0, \hskip4pt \mathrm{1\mkern-5muI\!}\hskip2pt +
\ast\hskip 2pt d_{\scriptscriptstyle 1})} \nonumber \\
&& \mkern-250mu
\frac {det(\phi^{\dagger}_{\scriptscriptstyle 0}
\phi^{\phantom{\dagger}}_{\scriptscriptstyle 0})^{\scriptscriptstyle 1/2}}
{det(\phi^{\dagger}_{\scriptscriptstyle 1}
\phi^{\phantom{\dagger}}_{\scriptscriptstyle 1})^{\scriptscriptstyle 1/2}}
det'(d^{\dagger}_{\scriptscriptstyle 0}
d^{\phantom{\dagger}}_{\scriptscriptstyle 0})^{\scriptscriptstyle 1/2}
det'(d^{\dagger}_{\scriptscriptstyle 1}
d^{\phantom{\dagger}}_{\scriptscriptstyle 1})^{\scriptscriptstyle -1/4}
Z_{\scriptscriptstyle SD}
\end{eqnarray}
We now note that expression (36) for the ratio of the partition functions 
of Maxwell--Chern--Simons theory and the Self--Dual Model is equal, 
modulo phase factor, to the partition function of pure abelian Chern--Simons  
theory\footnote[1]{This relationship, without the phase factor, was 
established in~\cite{arias} and \cite{steph} by different analyses. 
We would like to thank P. J. Arias and J. Stephany for drawing our attention 
to these references.}. Thus:
\begin{equation}
\frac{\raise 2pt \hbox{$ Z_{\scriptscriptstyle MCS}$}}
{\lower 2pt \hbox{$Z_{\scriptscriptstyle SD}$}} 
= e^{i \alpha} \hskip4pt Z_{\scriptscriptstyle CS}
\end{equation}
where
\begin{eqnarray}
\alpha = 
- \frac{\pi}{2} \zeta\biggl(0, -\Bigl(\ast d_{\scriptscriptstyle 1}
+ (\ast d_{\scriptscriptstyle 1})^{\scriptscriptstyle 2}\Bigr)_
{\scriptscriptstyle -}\biggr) 
&&
\mkern-28mu
+ \frac{\pi}{2}  
\zeta\Bigl(0, - (\ast d_{\scriptscriptstyle 1})_{\scriptscriptstyle -}\Bigr)
\!+\! \frac{\pi}{2} 
\zeta\Bigl(0, -(\mathrm{1\mkern-5muI}+\ast d_{\scriptscriptstyle 1})_
{\scriptscriptstyle -}\Bigr) 
\nonumber \\
&&\mkern-170mu
+ \frac{\pi}{4} \eta(0, \hskip4pt \ast \hskip2pt d_{\scriptscriptstyle 1})
+ \frac{\pi}{4} \eta(0, \hskip4pt \mathrm{1\mkern-5muI\!}\hskip2pt +
\ast\hskip 2pt d_{\scriptscriptstyle 1}) 
- \frac{\pi}{4} \eta\Bigl(0, \hskip4pt 
\ast d_{\scriptscriptstyle 1} +  
(\ast d_{\scriptscriptstyle 1})^{\scriptscriptstyle 2}\Bigr)
\end{eqnarray}
The partition function of abelian Chern--Simons 
theory~\cite{david},~\cite{s3} is equal, modulo phase factor, to the square 
root of the Ray-Singer torsion which is a topological invariant of the 
manifold given by~\cite{david}:
\begin{equation} 
\tau_{\scriptscriptstyle RS}(M)= \prod\limits_{\scriptscriptstyle q=0}^
{\scriptscriptstyle 3}\Bigl(\vert det\phi_{\scriptscriptstyle q}\vert \hskip 4pt
\vert det'd_{\scriptscriptstyle q}\vert \Bigr)^{\scriptscriptstyle (-1)^
{\scriptscriptstyle q}}
\end{equation}
So the absolute value of ratio of the partition functions of 
Maxwell--Chern--Simons theory and the Self--Dual Model is independent of the 
metric of the manifold and consequently these two theories are equivalent to 
within a phase factor on arbitrary manifolds. \\
Consider now the partition function of Maxwell--Chern--Simons theory 
with an external source $J$ coupled to the fields $A$:
\begin{equation}
Z_{\scriptscriptstyle MCS} (J) = \frac{1}{N_{\scriptscriptstyle MCS}}
\int\limits_{\Omega^{\scriptscriptstyle 1}(M)}
\mathcal{D}A \hskip 4pt e^{-i \hskip4pt \bigl\langle A, \hskip 4pt 
\left(\ast d_{\scriptscriptstyle 1} +
(\ast d_{\scriptscriptstyle 1})^{\scriptscriptstyle 2} \right)A \bigr\rangle
\hskip4pt + \hskip4pt \langle J, \hskip4pt A\rangle}
\end{equation}
For consistency we require that:
\begin{equation}
J \in ker\Bigl(\ast d_{\scriptscriptstyle 1} +
(\ast d_{\scriptscriptstyle 1})^{\scriptscriptstyle 2}\Bigr)^{\perp}
\end{equation}
Decompose $\Omega^{\scriptscriptstyle 1}(M)$:
\begin{equation}
\Omega^{\scriptscriptstyle 1}(M) = ker\Bigl(\ast d_{\scriptscriptstyle 1} +
(\ast d_{\scriptscriptstyle 1})^{\scriptscriptstyle 2}\Bigr) \oplus
ker\Bigl(\ast d_{\scriptscriptstyle 1} +
(\ast d_{\scriptscriptstyle 1})^{\scriptscriptstyle 2}\Bigr)^{\perp}
\end{equation}
Using (18) we may write:
\begin{eqnarray}
Z_{\scriptscriptstyle MCS} (J) = \frac{1}
{N_{\scriptscriptstyle MCS}} \hskip6pt
\mkern-30mu
&&vol\Bigl(ker(\ast d_{\scriptscriptstyle 1})\Bigr) \hskip4pt
vol\Bigl(ker(\mathrm{1\mkern-5muI\!} \hskip 2pt + 
\ast \hskip 2ptd_{\scriptscriptstyle 1})\Bigr) 
\nonumber \\
&&\int\limits_{ker\bigl(\ast d_{\scriptscriptstyle 1} +
(\ast d_{\scriptscriptstyle 1})^{\scriptscriptstyle 2}\bigr)^{\perp}}
\mkern-50mu
\mathcal{D}A \hskip 6pt
e^{- i \hskip4pt \bigl\langle A, \hskip 4pt 
\ast d_{\scriptscriptstyle 1} (\mathrm{1\mkern-5muI\!} +
\ast d_{\scriptscriptstyle 1})A \bigr\rangle
\hskip4pt + \hskip4pt \langle J, \hskip4pt A\rangle}
\end{eqnarray}
The integral gives:
\begin{eqnarray}
\int\limits_{ker\bigl(\ast d_{\scriptscriptstyle 1} +
(\ast d_{\scriptscriptstyle 1})^{\scriptscriptstyle 2}\bigr)^{\perp}}
\mkern-50mu
\mathcal{D}A \hskip5pt
e^{-i \hskip3pt \bigl\langle A, \hskip 4pt 
\ast d_{\scriptscriptstyle 1} (\mathrm {1\mkern-5muI\!} +
\ast d_{\scriptscriptstyle 1})A \bigr\rangle
\hskip4pt + \hskip4pt \langle J, \hskip4pt A\rangle} = \nonumber \\ 
&&\mkern-200mu
= det'\biggl(i \Bigl(\ast d_{\scriptscriptstyle 1} + 
(\ast d_{\scriptscriptstyle 1})^
{\scriptscriptstyle 2}\Bigr)\biggr)^{\scriptscriptstyle -1/2} \hskip 5pt
e^{i\hskip3pt \langle J, \hskip4pt \frac{1}{\ast d_{\scriptscriptstyle 1} + 
(\ast d_{\scriptscriptstyle 1})^{\scriptscriptstyle 2}} \hskip3pt J \rangle}
\end{eqnarray}
Using (30) we obtain:
\begin{eqnarray}
Z_{\scriptscriptstyle MCS} (J) = 
\frac{e^{ - \frac{i \pi}{2} \psi - \frac{i \pi}{4} 
\eta\bigl(0, \hskip4pt \ast d_{\scriptscriptstyle 1} +  
(\ast d_{\scriptscriptstyle 1})^{\scriptscriptstyle 2}\bigr)}}
{N_{\scriptscriptstyle MCS}} \hskip5pt
&& \mkern-30mu
vol\Bigl(ker(\ast d_{\scriptscriptstyle 1})\Bigr)
\hskip4pt
vol\Bigl(ker(\mathrm{1\mkern-5muI\!} \hskip 2pt + 
\ast \hskip 2ptd_{\scriptscriptstyle 1})\Bigr) \hskip 4pt \nonumber \\
&& \mkern-120mu
det'(\ast d_{\scriptscriptstyle 1})^{\scriptscriptstyle -1/2}
det'(\mathrm{1\mkern-5muI\!} + \ast d_{\scriptscriptstyle 1})^
{\scriptscriptstyle -1/2} 
e^{i\hskip3pt \langle J, \hskip4pt \frac{1}{\ast d_{\scriptscriptstyle 1} + 
(\ast d_{\scriptscriptstyle 1})^{\scriptscriptstyle 2}} \hskip3pt J \rangle}
\end{eqnarray}
Here we identify the Ray--Singer torsion~\cite{s4}. Namely, with 
suitable choice of normalization $N$:
\begin{equation}
\frac{1}{N} \hskip4pt
vol\Bigl(ker(\ast d_{\scriptscriptstyle 1})\Bigr) \hskip5pt
det'(\ast d_{\scriptscriptstyle 1})^{\scriptscriptstyle -1/2}
= \tau_{\scriptscriptstyle RS}(M)^{\scriptscriptstyle 1/2}
\end{equation}
Hence:
\begin{eqnarray}
Z_{\scriptscriptstyle MCS} (J) = \frac{ 
e^{- \frac{i \pi}{2} \psi
 - \frac{i \pi}{4} \eta\bigl(0, \hskip4pt \ast d_{\scriptscriptstyle 1} +  
(\ast d_{\scriptscriptstyle 1})^{\scriptscriptstyle 2}\bigr)}}
{N_{\scriptscriptstyle MCS}}  
det'(\mathrm{1\mkern-5muI\!} + \ast d_{\scriptscriptstyle 1})^
{\scriptscriptstyle -1/2} 
vol\Bigl(ker(\mathrm{1\mkern-5muI\!} \hskip 2pt + 
\ast \hskip 2ptd_{\scriptscriptstyle 1})\Bigr) \hskip 4pt 
\nonumber \\
&&
\mkern-240mu 
\tau_{\scriptscriptstyle RS}(M)^{\scriptscriptstyle 1/2}
e^{i \hskip3pt \langle J, \hskip4pt \frac{1}{\ast d_{\scriptscriptstyle 1} + 
(\ast d_{\scriptscriptstyle 1})^{\scriptscriptstyle 2}} \hskip3pt J \rangle}
\end{eqnarray}
The determinant entering this expression can be written as:
\begin{eqnarray}
det'(\mathrm{1\mkern-5muI\!} + \ast d_{\scriptscriptstyle 1})^
{\scriptscriptstyle -1/2} & = & 
e^{\frac{i \pi}{4} \eta (0,\hskip3pt \mathrm{1\mkern-5muI\!} + 
\ast d_{\scriptscriptstyle 1})}
det'\Bigl(- i \hskip3pt 
(\mathrm{1\mkern-5muI\!} + \ast d_{\scriptscriptstyle 1})\Bigr)^
{\scriptscriptstyle -1/2} 
\nonumber \\
&& \mkern-100mu
= e^{\frac{i \pi}{4} \eta (0, \hskip3pt \mathrm{1\mkern-5muI\!} + 
\ast d_{\scriptscriptstyle 1})}
e^{i \hskip3pt \langle J, \hskip4pt \frac{1}{\mathrm{1\mkern-5muI\!} \hskip 2pt + 
\ast d_{\scriptscriptstyle 1}} \hskip3pt J \rangle}
\mkern-35mu
\int\limits_{ker(\mathrm{1\mkern-5muI\!} \hskip 2pt + 
\ast \hskip 2ptd_{\scriptscriptstyle 1})^{\perp}}
\mkern-30mu
\mathcal{D}A \hskip 6pt
e^{-i \hskip3pt \bigl\langle A, \hskip 4pt 
- (\mathrm{1\mkern-5muI\!} + \ast d_{\scriptscriptstyle 1})A \bigr\rangle
\hskip4pt + \hskip4pt \langle J, \hskip4pt A\rangle}
\end{eqnarray}
In the integral we now change the variables from $A$ to $iA$. 
The Jacobian of this change of variables is 
$det'(i1\mathrm{\mkern-5muI})$ which is a constant and we can absorb it in the
normalization factor. \\
The product of this determinant with the volume element gives (modulo 
normalization factor) the partition function of the 
Self--Dual Model with current $\hat{J}=-iJ$. Therefore:
\begin{equation}
Z_{\scriptscriptstyle MCS} (J) = 
e^{i \alpha - \frac{i\pi}{4} 
\eta (0, \hskip 3pt \ast d_{\scriptscriptstyle 1})}
\tau_{\scriptscriptstyle RS}(M)^{\scriptscriptstyle 1/2}
e^{i \langle J, \hskip 3pt \frac{1}{\ast d_{\scriptscriptstyle 1}} \hskip4pt J \rangle}
Z_{\scriptscriptstyle SD} (\hat{J})
\end{equation}
The first exponent contains the geometrical information of the manifold via 
the $\eta$--function, while the second one yields the linking number of the 
currents. The partition function of pure abelian Chern--Simons 
theory~\cite{s4} in the presence of a current $J$ is
\begin{equation}
Z_{\scriptscriptstyle CS}(J) = e^{-\frac{i\pi}{4} 
\eta (0, \hskip 3pt \ast d_{\scriptscriptstyle 1})}
\tau_{\scriptscriptstyle RS}(M)^{\scriptscriptstyle 1/2}
e^{i \langle J, \hskip 3pt 
\frac{1}{\ast d_{\scriptscriptstyle 1}} \hskip4pt J \rangle}
\end{equation}
Therefore, at the level of currents, the ratio of the partition functions of 
Maxwell--Chern--Simons theory and Self--Dual Model is a topological 
invariant to within a phase factor:
\begin{equation}
\frac{Z_{\scriptscriptstyle MCS} (J)}{Z_{\scriptscriptstyle SD} (\hat{J})} =
e^{i \alpha}
\tau_{\scriptscriptstyle RS}(M)^{\scriptscriptstyle 1/2}
e^{i \langle J, \hskip 3pt 
\frac{1}{\ast d_{\scriptscriptstyle 1}} \hskip4pt J \rangle} =
e^{i \alpha} Z_{\scriptscriptstyle CS} (J) 
\end{equation}
where:
\begin{eqnarray}
\alpha = 
- \frac{\pi}{2} \zeta\biggl(0, -\Bigl(\ast d_{\scriptscriptstyle 1}
+ (\ast d_{\scriptscriptstyle 1})^{\scriptscriptstyle 2}\Bigr)_
{\scriptscriptstyle -}\biggr) 
&&
\mkern-28mu
+ \frac{\pi}{2}  
\zeta\Bigl(0, - (\ast d_{\scriptscriptstyle 1})_{\scriptscriptstyle -}\Bigr)
\!+\! \frac{\pi}{2} 
\zeta\Bigl(0, -(\mathrm{1\mkern-5muI}+\ast d_{\scriptscriptstyle 1})_
{\scriptscriptstyle -}\Bigr) 
\nonumber \\
&&\mkern-200mu
- \frac{\pi}{4} \eta\Bigl(0, \hskip4pt 
\ast d_{\scriptscriptstyle 1} +  
(\ast d_{\scriptscriptstyle 1})^{\scriptscriptstyle 2}\Bigr) +
\frac{\pi}{4} \eta(0, \hskip4pt \ast d_{\scriptscriptstyle 1}) 
+ \frac{\pi}{4} \eta(0, \hskip4pt \mathrm{1\mkern-5muI\!}\hskip2pt +
\ast\hskip 2pt d_{\scriptscriptstyle 1}) 
\end{eqnarray}
The factor $\alpha$ contains the geometrical information of 
the manifold.\\
Note that the correlation functions can be calculated in the usual way 
by functionally differentiating the partition functions with respect to the
external current. Equation (51) allows us to relate the correlation functions
between the models. \\
As an example, let us take the manifold to be $S^{\scriptscriptstyle 3}$ 
(hence the Ray--Singer torsion is 1~\cite{rs}) and
let us suppose that the currents do not link. Then we get 
$Z_{\scriptscriptstyle CS} (J) = 1$ and therefore Maxwell-Chern-Simons
theory is equivalent to the Self-Dual Model to within a phase factor which 
captures the geometrical properties of the manifold. If the currents link then
the partition functions differ by an additional phase which captures the 
topological features of the currents. \\
The main differences between our results and those of earlier authors can be 
summarized as follows. We consider arbitrary manifolds and show, for the complete
theories, the surprising result that the ratio of these two theories is itself a 
complete topological field theory (i.e. Chern--Simons theory). We also note that
when the manifold is I$\mkern-3mu$R$\mkern-10mu\phantom{s}^{\scriptscriptstyle 3} 
(S^{\scriptscriptstyle 3})$ and, as considered by earlier authors, with no 
topological entanglement of currents, then the partition function of the 
Chern--Simons theory is 1. This result is in exact agreement with the results obtained 
by earlier authors.

\section*{Acknowledgements} 
E. P. would like to thank David Adams for valuable discussions and 
Fabian Sievers for \LaTeX \hskip 4pt tips.

\end{document}